\documentclass{Interspeech2024}

\interspeechcameraready

\title{Qifusion-Net: Layer-adapted Stream/Non-stream Model for End-to-End Multi-Accent Speech Recognition}

\name[affiliation={1}]{Jinming}{Chen}
\name[affiliation={1}]{Jingyi}{Fang}
\name[affiliation={1}]{Yuanzhong}{Zheng}
\name[affiliation={1}]{Yaoxuan}{Wang}
\name[affiliation={1}]{Haojun}{Fei}

\address{$^1$Qifu Technology, China}
\email{chenjinming-jk@360shuke.com, fangjingyi-jk@360shuke.com, zhengyuanzhong-jk@360shuke.com, wangyaoxuan-jk@360shuke.com, zhangchulan-jk@360shuke.com}
\keywords{multi-accent speech recognition, layer-adapted fusion, stream/non-stream decoding, cross-attention}

\begin{document}

\maketitle

\begin{abstract}
    
    Currently, end-to-end (E2E) speech recognition methods have achieved promising performance. However, auto speech recognition (ASR) models still face challenges in recognizing multi-accent speech accurately. We propose a layer-adapted fusion (LAF) model, called Qifusion-Net, which does not require any prior knowledge about the target accent. Based on dynamic chunk strategy, our approach enables streaming decoding and can extract frame-level acoustic feature, facilitating fine-grained information fusion. Experiment results demonstrate that our proposed methods outperform the baseline with relative reductions of 22.1$\%$ and  17.2$\%$ in character error rate (CER) across multi accent test datasets on KeSpeech and MagicData-RMAC.
\end{abstract}

\section{Introduction}

In recent years, end-to-end (E2E) speech recognition (ASR) has significantly benefited from the high-resource languages and increasing large model size \cite{radford2023robust,gao2022paraformer}. This improvement has enabled the effective mitigation of recognition degraded caused by various acoustic environments \cite{chang2021exploration}. As a result, E2E ASR systems have found widely-used in commercial speech recognition products \cite{watanabe2018espnet,inaguma2020espnet,zhang2022wenet,gao2023funasr}. However, it is well known that the performance of even large ASR models degrades significantly when the speakers have varying degrees of accent pronunciation \cite{tang2021kespeech}. Accent is a special way of pronunciation, which is mainly influenced by regional culture, speaking style and the education level of the speaker \cite{Wang_Long_Li_Wei_2023}. For example, in the remote areas or villages of southern China, those accents are quite different from the pronunciation of Mandarin and seriously affects the recognition accuracy of the E2E ASR model \cite{tang2021kespeech}. To some extent, training a specific accent ASR model can solve the problem of recognition accuracy degradation. Achieving high recognition accuracy in multi-accent system without pre-accent category information has significant commercial value for the application of E2E ASR model.

Recently, adversarial learning \cite{na2021accented,hu2021redat}, transfer learning \cite{das2021best,luo2021cross}, multi-tasking learning (MTL) \cite{shao2023decoupling,dan2022multi} and other deep learning methods have been developed greatly to eliminate the recognition bias due to the accent in the ASR task. The primary concept behind adversarial learning and transfer learning methods involves initially training the model on an extensive corpus of speech data, followed by fine-tuning it specifically on the accent dataset \cite{sun2018domain,das2021best,maison2023improving}. Good performance is often obtained in a single accent system. For multi-accent systems, it is often necessary to introduce additional information to guide for the different accents. The direct way to introduce accent information is to concatenate a one-hot accent vector into the input acoustic features \cite{ai2020new}. Others, accent identification (AID) models are used to generate embeddings \cite{qian2022layer,ghorbani2023advanced}. For instance, the authors in \cite{Jain_Upreti_Jyothi_2018,Turan_Vincent_Jouvet_2020} suggest connecting accent embeddings and acoustic features to adapt the acoustic model. In \cite{Gong_Lu_Zhou_Qian_2021}, they utilized well-trained accent classifiers to extract accent embedding for layer-to-layer adaptation of E2E ASR models. A multi-task framework was proposed in \cite{Jain_Upreti_Jyothi_2018,Jain_Singh_Rath_2019} to jointly model ASR and AID tasks. All previous researches have significantly enhanced the accuracy of accent speech recognition in specific contexts. However, thus far, there has been a lack of consideration for accent recognition in real-time streaming scenarios and the fusion of fine-grained accent information at the frame level, which holds greater practical applicability and reference significance for MTL.

In this paper, we aim at improving the recognition accuracy in a multi-accent system. Three contributions are explored: 1) A new fusion strategy Layer-adapted fuison (LAF) module is proposed to extract accent information in shared acoustic encoder. 2) Fine-grained fusion of accent information at frame-level is achieved through the utilization of a cross-attention module, which effectively eliminates the impact of accent on the acoustic model. 3) Based on the dynamic chunk strategy, the model realizes the unification of streaming and non-streaming decoding modes. The results demonstrate that both stream/non-stream Qifusion-Net achieve the highest accuracy for accented speech datasets. The CER demonstrates the relative decrease of 22.1$\%$ and  17.2$\%$ in the multi accent test datasets, compared to the baseline in KeSpeech and MagicData-RAMC.

The remaining sections of this paper are structured as follows:  Section 2 introduces our Layer-adapted fusion module integrated with the MTL E2E ASR system. In Section 3, experimental results are presented and analyzed. Finally, Section 4 provides the conclusion.

\begin{figure*}[t]
	\centering
	\includegraphics[width=15cm]{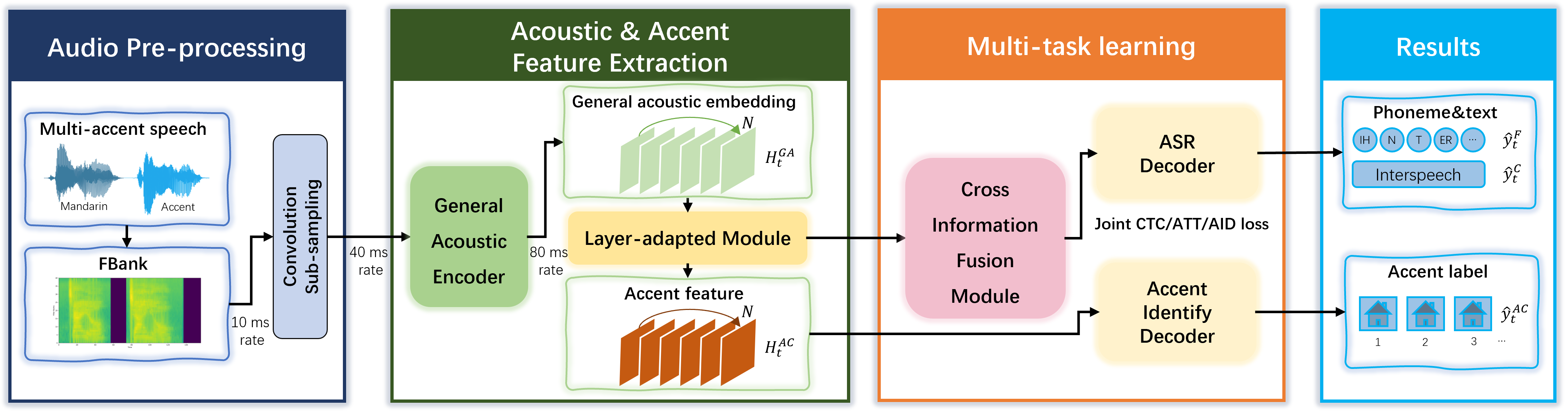}
	\caption{Schematic architecture of the proposed layer-adapted for end-to-end multi-accent ASR model.}
	\label{fig1}
\end{figure*}

\begin{figure*}[t]
	\centering
	\subfigure[General Acoustic Encoder]{
		\includegraphics[height=4cm]{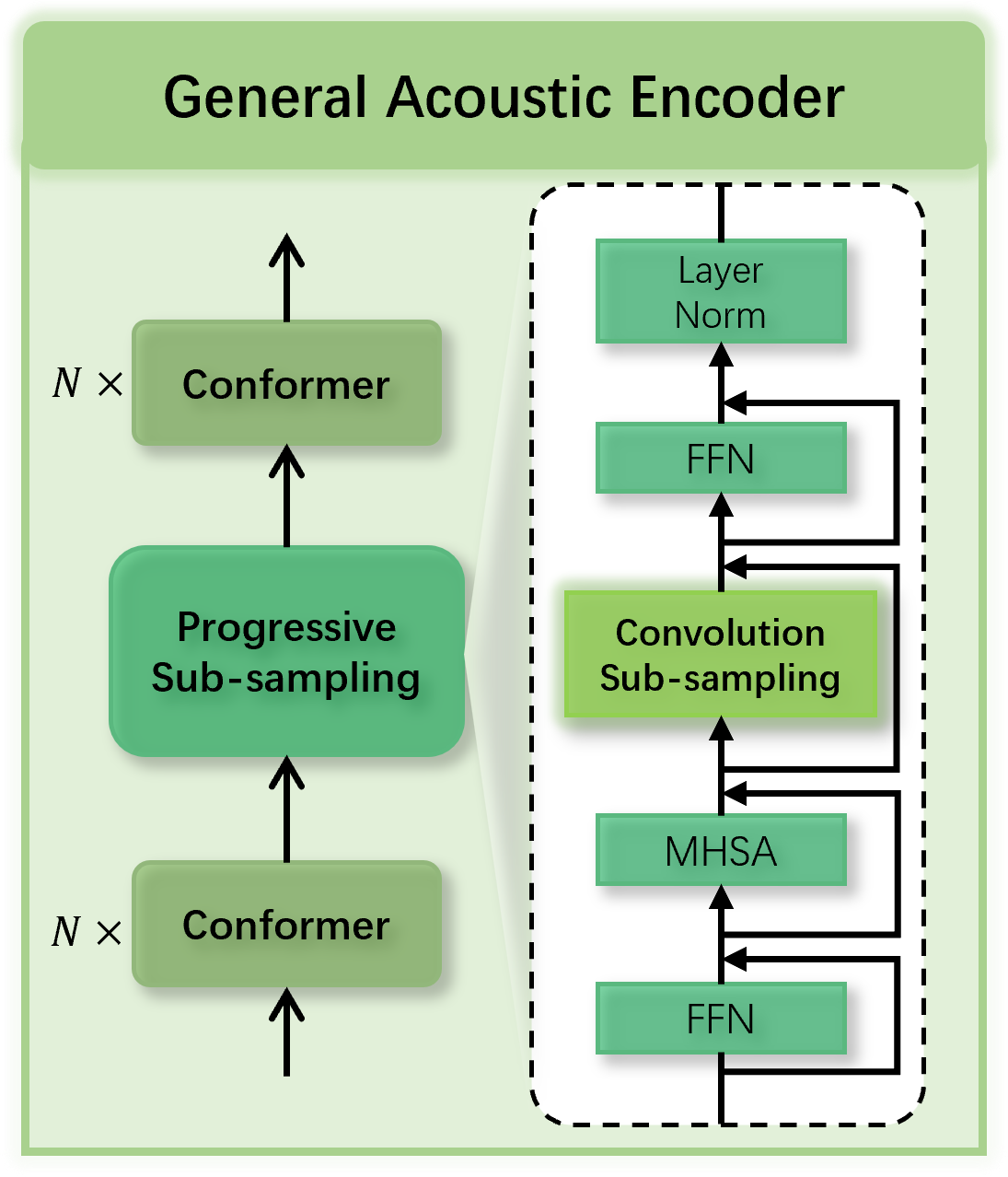}
		\label{fig2a}
	}
	\subfigure[Layer-adapted Module]{
		\includegraphics[height=4cm]{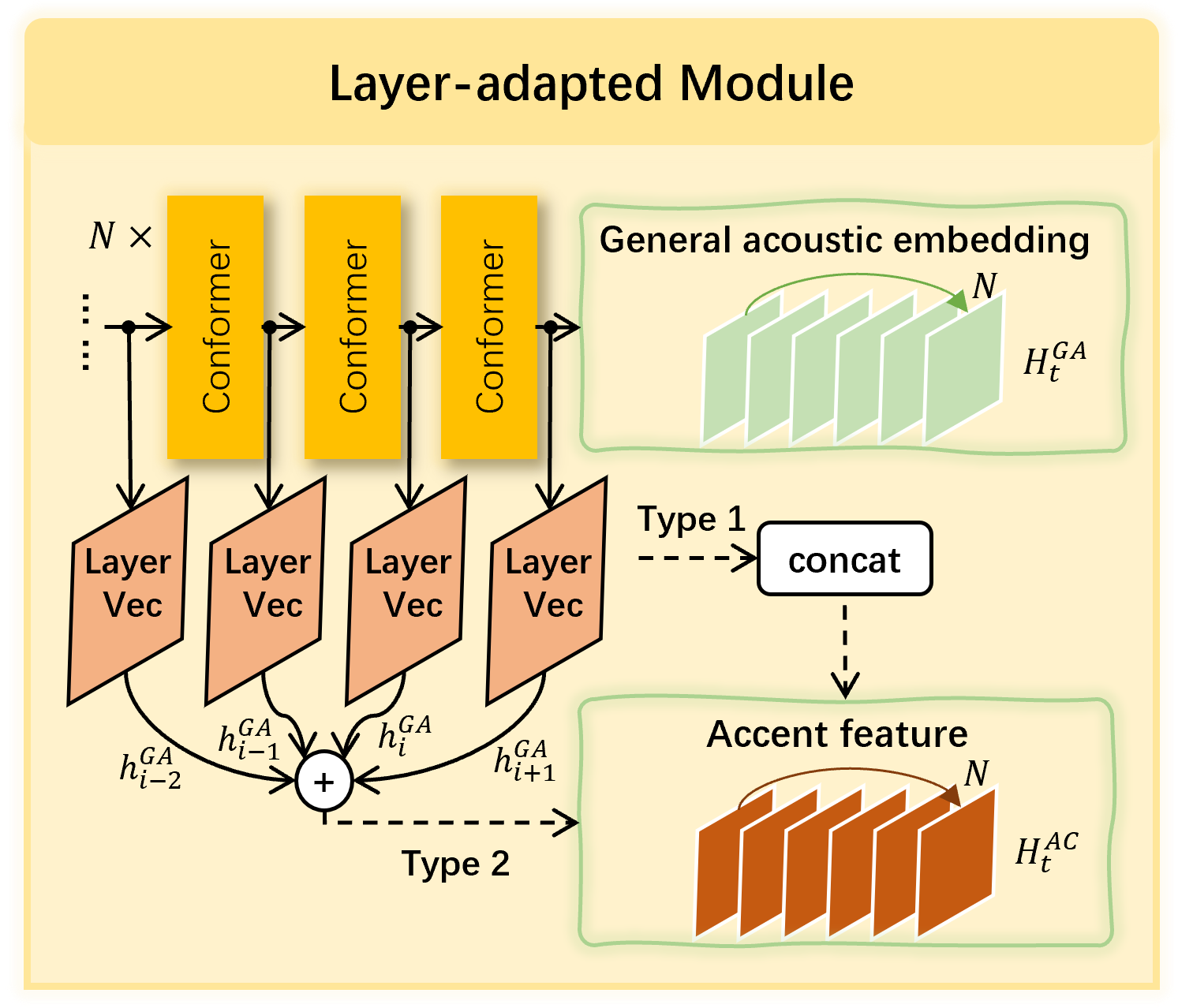}
		\label{fig2b}
	}
	\subfigure[Cross Information Fusion Module]{
		\includegraphics[height=4cm]{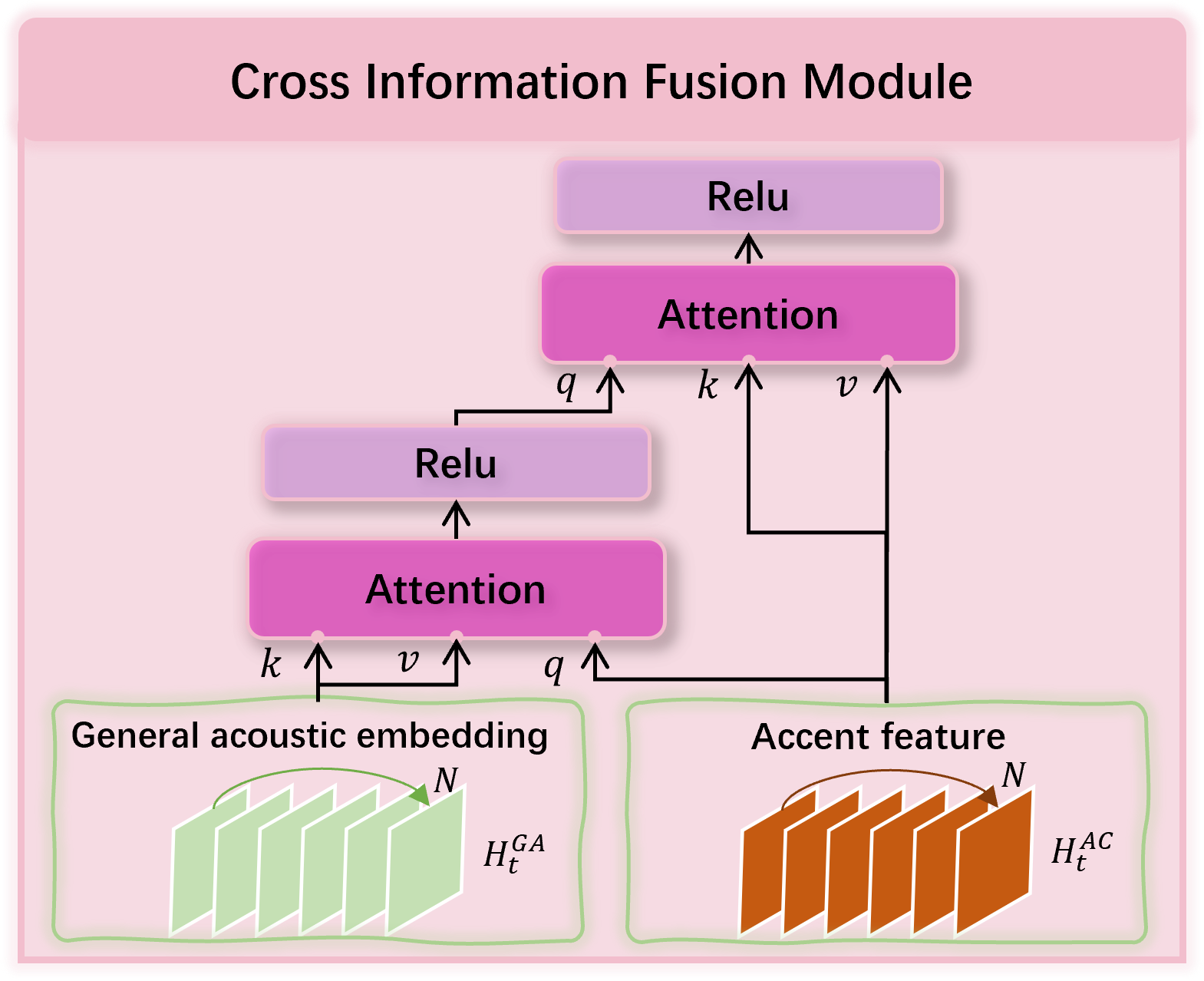}
		\label{fig2c}
	}
	\caption{Key parts of the model architecture.}
	\label{fig2}
\end{figure*}

\section{Layer-adapted for E2E Multi-Accent ASR}

The whole model architecture of the proposed layer-adapted for E2E multi-accent ASR is illustrated in \Cref{fig1}. In this section, we first give a brief review of a general acoustic encoder based on the convolution-augmented transformer (Conformer) model in Section 2.1. Then the proposed layer-adapted fusion (LAF) module is introduced in Section 2.2. The LAF module is designed for filtering the fine-grained accent information from the shared acoustic encoder with two different structures. Furthermore, we present the cross-attention module in Section 2.3, which integrates the acoustic and accent information. Finally, the MTL training of our methods and the stream/non-stream decoding modes are described in Section 2.4.

\subsection{Conformer-based Acoustic Encoder}

The acoustic encoder structure innovations in this paper are based on the general conformer E2E ASR model. The Conformer model integrates convolution layers into the transformer module to enhance the local modeling capability of signal sequences. Given its exceptional performance in the E2E ASR task, we employ the conformer-based encoder depicted in \Cref{fig2a} to comprehensively extract frame-level acoustic features from incoming multi-accent speech.

The original conformer model is mainly composed of four modules: two feed-forward modules (FFN), a multi-head self-attention module (MHSA) and a convolution module in the middle. When the step length of speech frame is set to 10ms, there will be redundancy in the general acoustic features which will affect the fine-grained accent feature extraction to some extent. Hence a progressive sub-sampling operation is applied in the general acoustic encoder. The feature compression in time dimension will be conducive to the following effective accent features extraction by layer-adapted module in Section 2.2. 

Given the FBank features $ X_t $ of the input multi-accent speech, the frame-level general acoustic features $ H_{t}^{GA} $ of the conformer block can be mathematically defined as follows:

\begin{equation}
	h_{t}^{\mathrm{FFN}_{1}} = X_t+\frac{\mathrm{FFN} (X_t)}{2} 
\end{equation}
\begin{equation}
	h_{t}^{\mathrm{MHSA}} =h_{t}^{\mathrm{FFN} _{1}} +{\mathrm{MHSA} (h_{t}^{\mathrm{FFN} _{1}})}
\end{equation}
\begin{equation}
	h_{t}^{\mathrm{Conv} } = h_{t}^{\mathrm{MHSA}} + {\mathrm{Conv} (h_{t}^{\mathrm{MHSA}})}
\end{equation}
\begin{equation}
	h_{t}^{\mathrm{FFN} _{2}} = h_{t}^{\mathrm{Conv} } + \frac{{\mathrm{FFN} (h_{t}^{\mathrm{Conv} })}}{2} 
\end{equation}
\begin{equation}
	H_{t}^{GA} = {{\mathrm{LayerNorm} (h_{t}^{\mathrm{FFN} _{2}})}}
\end{equation}

 On the one hand, this design can gradually reduce the dimension of input sequence, so that the global acoustic features can be projected to a wider dimension \cite{Burchi_Vielzeuf_2021}, and on the other hand, it can reduce the computational complexity of streaming mode.

\subsection{Layer-adapted Module}

Current solutions for multi-accent ASR task commonly adopt the acoustic features of a universal ASR model as the input to train an accent identifier (AID) model. However, such methods lack information sharing between the two models which leads to both performance degradation in a multi-task learning of ASR and AID. As shown in \Cref{fig2b}, we proposed a layer-adapted module to extract fine-grained accent information from different layers of acoustic encoder while facilitating frame-by-frame correction of ASR results by cross-attention module in Section 2.3.

\begin{table}[]
	\caption{Ablation and contrast experiments results of the AID task on the KeSpeech test dataset.}
	\label{tab:aid}
	\renewcommand\arraystretch{1}
	\setlength{\tabcolsep}{7mm}{
		\begin{tabular}{ccc}
			\hline \toprule[0.3pt]
			&                                  & \textbf{AID}                          \\
			\multirow{-2}{*}{\textbf{ID}} & \multirow{-2}{*}{\textbf{Model}} & \textbf{ACC(\%)}                      \\ \hline \toprule[0.3pt]
			C1                            & KeSpeech Baseline                & 61.13                                 \\
			C2                            & Kaldi-xvector                    & 56.34                                 \\
			C3                            & ResNet-34                        & {\color[HTML]{000000} 61.13}          \\
			C4                            & ECAPA-TDNN                       & {\color[HTML]{000000} 60.77}          \\
			C6                            & DIMNet w/ LM                     & {\color[HTML]{000000} {\ul 78.57}}    \\ \hline
			L1                            & Qifusion-Net-L6                  & {\color[HTML]{000000} 33.64}          \\
			L2                            & Qifusion-Net-L7                  & {\color[HTML]{000000} 34.67}          \\
			L3                            & Qifusion-Net-L8                  & {\color[HTML]{000000} \textbf{39.75}} \\
			L4                            & Qifusion-Net-L9                  & {\color[HTML]{000000} 33.08}          \\
			L5                            & Qifusion-Net-L10                 & {\color[HTML]{000000} 25.95}          \\
			L6                            & Qifusion-Net-L11                 & {\color[HTML]{000000} 26.17}          \\ \hline
			Q2                            & Qifusion-Net-ns                  & {\color[HTML]{000000} \textbf{79.10}} \\ \hline \toprule[0.3pt]
	\end{tabular} }
\end{table}


\begin{table*}[h] 
	\caption{Ablation and contrast experiments results of the proposed stream/non-stream Qifusion-Net on the KeSpeech dataset.}
	\label{tab:asr}
	\renewcommand\arraystretch{1}
	\centering
	\setlength{\tabcolsep}{1.5mm}{
		\begin{tabular}{ccccccccccccc}
			\hline \toprule[0.3pt]
			&                                                                   & \textbf{AID}                          & \multicolumn{10}{c}{\textbf{ASR WER(\%)}}                                                                                                                                                                                                                                                                                                                                                                                                                                \\
			\multirow{-2}{*}{\textbf{ID}} & \multirow{-2}{*}{\textbf{Model}}                                  & \textbf{ACC(\%)}                      & \textbf{Total} & \textbf{Beijing} & \textbf{Ji-Lu} & \textbf{\begin{tabular}[c]{@{}c@{}}Jiang\\ Huai\end{tabular}} & \textbf{\begin{tabular}[c]{@{}c@{}}Jiao\\ Liao\end{tabular}} & \textbf{\begin{tabular}[c]{@{}c@{}}Lan\\ Yin\end{tabular}} & \textbf{Mandarin} & \textbf{\begin{tabular}[c]{@{}c@{}}North\\ eastern\end{tabular}} & \textbf{\begin{tabular}[c]{@{}c@{}}South\\ western\end{tabular}} & \textbf{\begin{tabular}[c]{@{}c@{}}Zhong\\ yuan\end{tabular}} \\ \hline
			C1                            & \begin{tabular}[c]{@{}c@{}}KeSpeech\\ Baseline\end{tabular}       & 61.13                                 & 10.38          & 11.5             & 11.5           & 15.9                                                          & 11.7                                                         & 11.7                                                       & 6.1               & 10.2                                                             & 11.9                                                             & 9.6                                                           \\
			\multicolumn{1}{l}{C5}        & \multicolumn{1}{l}{DIMNet w/o LM}                                 & 78.57                                 & 9.40           & -                & -              & -                                                             & -                                                            & -                                                          & -                 & -                                                                & -                                                                & -                                                             \\
			C6                            & DIMNet w/ LM                                                      & {\color[HTML]{000000} {\ul 78.57}}    & 8.87           & -                & -              & -                                                             & -                                                            & -                                                          & -                 & -                                                                & -                                                                & -                                                             \\ \hline
			A1                            & \begin{tabular}[c]{@{}c@{}}Qifusion-Net\\ w/o lam\end{tabular}    & -                                     & 15.35          & 17.13            & 17.59          & 23.63                                                         & 17.93                                                        & 19.43                                                      & 8.32              & 11.69                                                            & 16.87                                                            & 14.19                                                         \\
			A2                            & \begin{tabular}[c]{@{}c@{}}Qifusion-Net\\ fusion-sum\end{tabular} & 73.41                                 & 9.03           & 10.93            & 10.4           & 14.41                                                         & 10.84                                                        & 10.12                                                      & 4.75              & 8.99                                                             & 10.68                                                            & 7.81                                                          \\
			A3                            & \begin{tabular}[c]{@{}c@{}}Qifusion-Net\\ w/o cif\end{tabular}    & 76.46                                 & 9.6            & 11.25            & 11.11          & 15.13                                                         & 11.24                                                        & 10.89                                                      & 5.16              & 8.99                                                             & 11.23                                                            & 8.46                                                          \\
			A4                            & \begin{tabular}[c]{@{}c@{}}Qifusion-Net\\ self-att\end{tabular}   & \textbf{79.10}                        & {\ul 8.25}     & {\ul 9.32}       & {\ul 9.37}     & {\ul 13.22}                                                   & {\ul 9.92}                                                   & {\ul 9.32}                                                 & {\ul 4.57}        & \textbf{7.77}                                                    & {\ul 9.28}                                                       & {\ul 7.37}                                                    \\ \hline
			Q1                            & Qifusion-Net-s                                                    & 76.54                                 & 8.9            & 10.43            & 10.18          & 14.33                                                         & 10.49                                                        & 10.08                                                      & 4.75              & 8.76                                                             & 10.24                                                            & 7.85                                                          \\
			Q2                            & Qifusion-Net-ns                                                   & {\color[HTML]{000000} \textbf{79.10}} & \textbf{8.08}  & \textbf{9.71}    & \textbf{9.15}  & \textbf{12.8}                                                 & \textbf{9.58}                                                & \textbf{9.18}                                              & \textbf{4.45}     & {\ul 8.28}                                                       & \textbf{9.27}                                                    & \textbf{7.12}                                                 \\ \hline \toprule[0.3pt]
	\end{tabular} }
\end{table*}

\subsubsection{Adapted Layer}

Numerous studies have indicated that distinct layers of the ASR encoder possess the capability to extract speech information at varying levels. As the depth of the ASR layers increases, a greater abundance of localized information becomes available. In this paper, we use the layer after the progressive sub-sampling operation of acoustic structure as adapted layers. In the training process, we introduce learnable adaptive weights that are multiplied with the adapted layers. Both concatenate and sum operations can be selected as layer-adapted connectivity options (\Cref{fig2b}).

\subsubsection{Accent Identify Decoder}

After extracting the fused accent features from the adapted layers in the acoustic encoder, we propose a two-layer causal convolutional structure and a linear-based discriminator to construct the AID. This module can effectively distill accent information and provides frame-by-frame classification of input multi-accent speech into different accent categories.

\subsection{Cross-attention Module}

It is widely acknowledged that the MTL approach can enhance the performance of each task by facilitating the sharing of feature information. As shown in \Cref{fig2c}, we use the output $ H_{t}^{GA} $ of the general acoustic encoder as the key, and the accent embedding $ H_{t}^{AC} $ obtained in the layer adapted module as the query to carry out cross-information fusion. The frame-level accent embedding features contributes to eliminate the distortion of acoustic features caused by different degrees of accent pronunciation in a multi-accent speech in order to improve the accuracy of accent ASR.

The following shows calculation process of cross-information fusion based on attention mechanism:

\begin{equation}
	Q_{t}  =W_{t}^{Q} H_{t}^{AC}, K_{t}=W_{t}^{K}H_{t}^{GA}, V_{t}=W_{t}^{V}H_{t}^{GA}
\end{equation}
\begin{equation}
	Q_{t}^{Att} = \mathrm{Relu} (\mathrm{Softmax} (\frac{Q_{t}(K_{t})_{}^{T}}{\sqrt{d_{att}} } V_{t})
\end{equation}
\begin{equation}
	O_{t}^{Att} = \mathrm{Relu} (\mathrm{Softmax} (\frac{Q_{t}^{Att}(K_{t})_{}^{T}}{\sqrt{d_{att}} } V_{t})
\end{equation}
where $ W_{t}^{Q} $, $ W_{t}^{K} $ and $ W_{t}^{V} $ are trainable weight matrix, the division of the similarity matrix and $  \sqrt{d_{att}} $ in (7) and (8) contribute to steady gradient descent while training.
\subsection{Multi-task Training and Stream/non-stream Decoding}

During the multi-task accent ASR model training, the overall loss function is designed by combining three losses: connectionist temporal classification (CTC) loss from ASR task, decoder attention loss and the accent identify cross entropy (CE) loss, which can be formulated as:

\begin{equation}
	\mathcal{L}_{all}=\mathcal{L}_{att}+\lambda_{ctc} \mathcal{L}_{ctc}+\lambda_{aid} \mathcal{L}_{aid}
\end{equation}
\begin{equation}
	\mathcal{L}_{ctc}=\mathrm{CTC} ({O_{t}^{Att}},y_{f})
\end{equation}
\begin{equation}
	\mathcal{L}_{att}=\mathrm{CE} (\mathrm{Decoder}(O_{t}^{Att},y_{c}) ,y_{c})
\end{equation}
\begin{equation}
	\mathcal{L}_{aid}=\mathrm{CE} (\hat{y} _{t}^{ac},y_{ac})
\end{equation}
where $ \lambda_{ctc} $ and $ \lambda_{aid} $ are two weights of CTC loss and AID loss. $ \hat{y} _{t}^{ac} $ and $ y_{ac} $ are the predicted and true accent label of the input $ X_{t} $. $ y_{c} $ and $ y_{f} $ are the transcription labels with coarse-grained and fine-grained units. $ \mathrm{CE}(\cdot ) $ and $ \mathrm{Decoder}(\cdot ) $ stands for the cross entropy loss and attention decoder function.

In this paper, we use the dynamic chunk masking strategy \cite{zhang2022wenet} to ensure compatibility of model inference with both stream and non-stream modes. During the training stage, we initially sample a random chunk size C from a uniform distribution ranging between 1 and the maximum batch length $ T $. Subsequently, the input is divided into multiple chunks based on the selected chunk size. Finally, in training, the current chunk undergoes bidirectional chunk-level attention with itself and previous/following chunks through left-to-right and right-to-left attention decoder respectively.

\section{Experiments}
\subsection{Dataset Description}

In this study, we conducted extensive experiments on KeSpeech \cite{tang2021kespeech}, which involved 1,542 hours of speech signals recorded by 27,237 speakers from 34 cities across China. The dataset encompasses standard Mandarin and its eight subdialects in the regions of Zhongyuan, Southwestern, Ji-Lu, Jiang-Huai, Lan-Yin, Jiao-Liao, Northeastern and Beijing. The MagicData-RAMC, which contains 180 hours and 6 diverse domains (Sichuan, Shanxi, Shandong, Jiangsu, Hunan, Guangdong), is also used as performance validation \cite{Yang_Chen_Luo_Yang_Ye_Cheng_Xu_Jin_Zhang_Zhang_et}.

\subsection{E2E Based Basedline}

For \cite{Wang_Long_Li_Wei_2023,Rabiner89-ATO} acoustic feature extraction, the 80-dimensional log Mel-filter bank (FBANK) is calculated with window size of 25ms and step size of 10ms. The utterance-level cepstral mean and variance normalization (CMVN) calculated using the training set was applied to FBANK for feature normalization. All our experiments are implemented using Wenet end-to-end speech processing toolkit. SpecAugment \cite{Park_Chan_Zhang_Chiu_Zoph_Cubuk_Le_2019} is used for data augmentation during training, and no extra language models are applied.

\subsection{Layer-adapted Fusion Module}
\subsubsection{layer-adapted Module}
As show in \Cref{fig1}, The Frame-level general acoustic embedding is obtain from the general acoustic encoder ($ h_{i}^{GA} $,  i stands for the index of layer). The fusion input ($ H_{t}^{GA} $) is stacked by $\left \{ h_{i}^{GA} ... h_{i+6}^{GA} \right \}  $. In our work, we chose the output of L-6th to L-12th after the progressive downsampling operation as the input for our layer-adapted module. Learnable weights $ W $ are introduced into each layer for dot multiplication. The input $W \times H_{t}^{GA} $ is fused through casual conv2d (kernal size 5x5, stride size 1x1). Accent frame-level predict label is calculated through conv1d of kenerl size 3 and stride size 1.

For AID task, we do some experiments to prove that the different layers of general acoustic encoder contain different-level accent information. We employ a pretrained acoustic encoder, freeze the encoder weights and utilize various layers as inputs to the AID model. \Cref{tab:aid} shows, without using layer-adapted module, L-8th has the highest accuracy for AID on the test set. From L-6th to L-8th, the accuracy increases as the layers deepen, but it declines to a certain extent after L-8th. The accuracy of the AID task with the layer-adapted module reaches 79.1$\%$, which exceeds the absolute improvements of both the KeSpeech baseline \cite{tang2021kespeech} (17.91$\%$) and DIMNet \cite{shao2023decoupling} (0.28$\%$).

We also investigated the performance differences between concatenate and weight sum connections in layer-adapted architectures. As shown in \Cref{tab:asr}, when compared to concatenate (Q2), weight sum (A2) exhibited a 10.5$\%$ relative increase in CER performance."

\subsubsection{Cross-attention Module}
Before the linear classification layer in AID task, we obtain the frame-level accent embedding ($ H_{t}^{AC} $). The ($ H_{t}^{AC} $) and  the last-layer of general acoustic encoder ($ h_{12}^{GA} $) have the same temporal resolution and feature dimension 256. $ Q_{t}^{Att} = CrossAtt(H_{t}^{AC}, h_{12}^{GA}) $. $ Q_{t}^{Att} $ is used as the attention decoder input with the accent bias eliminated. The attention decoder uses a bi-directional 3-layer transformer structure with a multi-head of 4.

For the AID-ASR MTL system, we do ablation experiments to discuss the impact of different modules on the overall CER. Without AID task, the overall CER reached 15.35 (A1). Added AID task, but without cross-attention module, CER increased to 9.6 (A3). Joint training has a significant positive effect on CER of ASR task. For accent information, cross-attention (Q2) achieves an absolute CER improvement of 0.17 compared to self-attention module (A4). It shows that the addition of cross-attention module can effectively eliminate the degradation of recognition caused by accent in ASR task.

\subsubsection{Stream/non-stream Decoding}
Based on the dynamic chunk masking strategy, the model supports steaming decoding mode. Compared with baseline and the sota model DIMNet, without LM, the CER of non-stream model (Qifusion-Net-ns Q2) is 10$\%$ higher than DIMNet without LM (C5) and even exceeds the DIMNet with LM (C6). The stream decoding mode (Qifusion-Net-s Q1) also outperforms C5 in terms of CER, reaching 8.9. This enables Qifusion-Net-s to match the recognition accuracy of sota model in streaming multi accent system scenarios, which has great potential in practical applications.

\Cref{tab:MagicData} presents the experimental result on the Magicdata-RMAC.  For the ASR task, the first row represents the official baseline \cite{Yang_Chen_Luo_Yang_Ye_Cheng_Xu_Jin_Zhang_Zhang_et}, while the second row corresponds to the conformer frameworks baseline \cite{gulati2020conformer}. The third row displays the latest CER results without a cross-modal extractor \cite{Wei_Li_Lv_Lu_Jiang_Xie_2023}. Finally, our proposed Qifusion-Net-ns achieves outstanding performance with a 17.2$\%$ lower CER compared to the baseline and attains an accuracy of 82.8$\%$ in the AID task.

\begin{table}[]
	\caption{Comparison the CER of different models on the MagicData dataset}
	\centering
	\label{tab:MagicData}
	\begin{tabular}{ccc}
		\hline
		\textbf{Model}                    & \textbf{\begin{tabular}[c]{@{}c@{}}AID \\ ACC(\%)\end{tabular}} & \textbf{\begin{tabular}[c]{@{}c@{}}RMAC/test\\  CER(\%)\end{tabular}} \\ \hline
		LAS-Conformer \cite{Yang_Chen_Luo_Yang_Ye_Cheng_Xu_Jin_Zhang_Zhang_et}                     & -                                                               & 19.1                                                                  \\
		\multicolumn{1}{l}{Conformer-ASR \cite{gulati2020conformer}} & -                                                               & 18.6                                                                  \\
		CVAE TT \cite{Wei_Li_Lv_Lu_Jiang_Xie_2023}                           & {\color[HTML]{000000} -}                                        & 17.6                                                                  \\
		Qifusion-net-ns                   & {\color[HTML]{000000} \textbf{82.8}}                            & \textbf{16.3}                                                         \\ \hline
	\end{tabular}
\end{table}

\section{Conclusion}
In this study, we explore an end-to-end asr decoding framework in multi-accent systems without prior accent information. Based on the standard conformer ASR architecture, we propose a Qifusion-Net, AID-ASR multi-task learning method based on shared progressive sub-sampling conformer encoder and layer-adapted fusion. We prove that layer-adapted improves AID task, while cross-fusion is beneficial for ASR tasks. The proposed method has lower CER than baseline and sota results in multi accent datasets.

\bibliographystyle{IEEEtran}
\bibliography{mybib}

\end{document}